\documentclass[3p]{elsarticle}
\usepackage{aasmacros}
\usepackage{amsmath, amssymb}
\usepackage{multicol}
\usepackage{xcolor}
\usepackage{xspace}
\usepackage{enumerate}
\usepackage{tablefootnote}
\usepackage{threeparttable}
\usepackage{booktabs}
\usepackage[english]{babel}
\usepackage[section]{placeins}
\usepackage[pdfpagelabels, plainpages=false, linktocpage=true]{hyperref}

\hypersetup{
   colorlinks   = true, 
   urlcolor     = blue, 
   linkcolor    = MidnightBlue, 
   citecolor    = red 
}

\journal{PoS: ICRC 2021}

\makeatletter
\let\oldtheequation\theequation
\renewcommand\tagform@[1]{\maketag@@@{\ignorespaces#1\unskip\@@italiccorr}}
\renewcommand\theequation{(\oldtheequation)}

\makeatother
\addto\extrasenglish{%
}

\begin{document}

\title{FR-0 jetted active galaxies: extending the zoo of candidate sites for UHECR acceleration}

\author[1]{Lukas Merten\corref{cor1}}
\ead{lukas.merten@uibk.ac.at}
\author[1]{Margot Boughelilba}
\ead{margot.boughelilba@uibk.ac.at}
\author[1]{Anita Reimer}
\ead{anita.reimer@uibk.ac.at}
\author[1]{Paolo Da Vela}
\ead{paolo.da-vela@uibk.ac.at}
\author[2]{Serguei Vorobiov}
\ead{serguei.vorobiov@gmail.com}
\author[3]{Fabrizio Tavecchio}
\ead{fabrizio.tavecchio@inaf.it}
\author[4]{Giacomo Bonnoli}
\ead{bonnoli@iaa.es}
\author[2]{Jon Paul Lundquist}
\ead{jplundquist@gmail.com}
\author[3]{Chiara Righi}
\ead{chiara.righi@inaf.it}

\address[1]{Institute for Astro and Particle Physics, University of Innsbruck, Technikerstraße 25, 6020 Innsbruck, Austria}
\address[2]{Center for Astrophysics and Cosmology (CAC), University of Nova Gorica,  Vipavska 13, SI-5000 Nova Gorica, Slovenia}
\address[3]{Astronomical Observatory of Brera, Via Brera 28, 20121 Milano, Italy}
\address[4]{Instituto de Astrofísica de Andalucía (CSIC), Apartado 3004, E-18080 Granada, Spain }

\cortext[cor1]{Corresponding author}

\begin{abstract}
Fanaroff-Riley (FR) 0 radio galaxies form a low-luminosity extension to the well-established ultra-high-energy cosmic-ray (UHECR) candidate accelerators FR-1 and FR-2 galaxies. Their much higher number density --- up to a factor five times more numerous than FR-1 with $z\leq 0.05$ --- makes them good candidate sources for an isotropic contribution to the observed UHECR flux. Here, the acceleration and survival of UHECR in prevailing conditions of the FR-0 environment are discussed.

First, an average spectral energy distribution (SED) is compiled based on the \textit{FR0CAT}. These photon fields, composed of a jet and a host galaxy component, form a minimal target photon field for the UHECR, which will suffer from electromagnetic pair production, photo-disintegration, photo-meson production losses, and synchrotron radiation. The two most promising acceleration scenarios based on Fermi-I order and gradual shear acceleration are discussed as well as different escape scenarios.

When an efficient acceleration mechanism precedes gradual shear acceleration, e.g., Fermi-I or others, FR-0 galaxies are likely UHECR accelerators. Gradual shear acceleration requires a jet Lorentz factor of Gamma>1.6, to be faster than the corresponding escape. In less optimistic models, a contribution to the cosmic-ray flux between the knee and ankle is expected to be relatively independent of the realized turbulence and acceleration.
\end{abstract}


\maketitle

\section{Introduction} 
\label{sec:intro}
The UHECRs flux, despite the observation of, e.g., a dipole in Auger data, is quite isotropic and can be explained in two ways: 1) UHECRs from a few luminous sources are isotropized during propagation in the intergalactic magnetic field (IGMF) or 2) many less luminous sources contribute to the flux, lowering the requirements on IGMF deflection. The latter ansatz has received more attention in recent years, e.g., in work on starburst galaxy contributions to the  UHECR flux (see, e.g., \cite{starburstAuger}).

The relatively new class of low-luminous Fanaroff-Riley 0 radio galaxies (FR-0s) (see, e.g., \cite{Baldi2009, Baldi2018}) could, in principle, contribute to the second scenario. They are about five times more numerous compared with their more luminous relatives, the well-established UHECR source class of FR~1 and FR~2 radio galaxies. Furthermore, their average jet luminosity of about $L_\mathrm{jet}\approx 10^{42}-10^{43.5}\,\mathrm{erg}\,\mathrm{s}^{-1}$ is well above the energetically required mean luminosity of $\sim 10^{40.5}\,\mathrm{erg}\,\mathrm{s}^{-1}$ to explain the UHECR energy density (see \cite{Merten2021} for details). 

To contribute to the observed UHECR flux, FR-0s must  accelerate particles --- including heavy nuclei up to iron --- to energies above $\sim 100$~EeV. To examine their potential to reach these energies, we take a sample of 114 observed FR-0 galaxies from the \textit{FR0CAT} to estimate their source environment. These estimates include descriptions of the target photon fields, the magnetic field strength, emission region size, and the jet's Lorentz factor. These parameters are combined to derive time scales for acceleration, escape, and losses (see Section \ref{sec:timescales}) and finally determine the maximal energy.

\section{Acceleration, Escape, and Losses}
\label{sec:timescales}

\subsection{Acceleration}
\label{ssec:acc}
The nature of cosmic-ray acceleration in FR-0 galaxies is unclear, and the knowledge about the exact condition in the source region is quite limited. Therefore, we pick two promising acceleration mechanisms, namely Fermi-1 order and gradual shear acceleration. Observations and modeling of FR-1/2 galaxies have shown that some type of shock acceleration is likely taking place in their jets. The similarities between the different types of FR galaxies justifies the assumption of the first mechanism. Furthermore, improved observations have revealed deviations between simple one-zone models and the data. A possible solution is to extend the source region model to a structured jet by including a velocity profile gradient (see, e.g., \cite{Ghisellini2005, Tavecchio2021}). This extension allows (gradual) shear acceleration to some extent.

\paragraph{Shock Acceleration:}
The acceleration time scale for Fermi-1 order acceleration is
\begin{align}
    \tau_\mathrm{acc}^\mathrm{Fermi} = \frac{20}{\alpha u_\mathrm{s}^2}\kappa (E)\quad ,
\end{align}
where $\kappa(E)\propto E^\alpha$ is the diffusion coefficient with spectral index $\alpha$ and $u_\mathrm{s}$ is the shock velocity. Here, we assume a strong shock $u_\mathrm{u}=4u_\mathrm{d}$ with a constant diffusion coefficient $\kappa_\mathrm{u}=\kappa_\mathrm{d}$, where quantities with subscript $d$ and $u$ are in the down- and upstream regime, respectively.

\paragraph{Gradual Shear Acceleration:}
Gradual shear acceleration is driven by consecutive scatterings in a flow with a velocity gradient. In contrast to \textit{shear} acceleration (see, e.g., \cite{Ostrowski1990}), the width of the shear profile is larger than the particle's gyroradius. Rieger and Duffy have shown that the acceleration time scale is given by \cite{Rieger2019, rieger_review}:
\begin{align}
     \tau_\mathrm{acc}(E) = \frac{5}{4+\alpha} \left( \frac{\partial u}{\partial r} \right)^{-2}\frac{c^2}{\Gamma_\mathrm{j}^4}\,\kappa(E)^{-1} \quad .
\end{align}
In this work, we assumed a linear flow profile $\partial u/\partial r=\Delta u/\Delta r$ over the full width of the jet. Here, the flow speed at the jet's edge vanishes ($u(r_\mathrm{em})=0$) and the central speed $u_\mathrm{j}\equiv u(r=0)=c\sqrt{1-\Gamma_\mathrm{j}^{-2}}$ is given by the estimated jet Lorentz factor $\Gamma_\mathrm{j}$.

\subsection{Escape}
\label{ssec:escape}
In this work, we take into account the two most relevant escape mechanisms --- advection and diffusion.

\paragraph{Advection:}
The escape time scale due to advection is given by:
\begin{align}
    \tau_\mathrm{esc}^\mathrm{adv} = \frac{L}{v_\mathrm{adv}}\quad ,
\end{align}
where $L$ is the relevant length scale and $v_\mathrm{adv}$ is the advection speed. Here, the length scale is given by the jet extension, which we assume to be at least $L\geq r_\mathrm{em}$ and the advection speed is given by the jet speed $v_\mathrm{adv}=u_\mathrm{j}$

\paragraph{Diffusion:}
Escape by diffusion is a quite common scenario, especially for gradual shear acceleration. The characteristic diffusion escape time is given by
\begin{align}
    \tau_\mathrm{esc}^\mathrm{dif} = \frac{L^2}{\kappa} \quad .
\end{align}
For the jet geometry discussed here, perpendicular diffusion with respect to the jet axis is the most efficient; therefore, the length scale is assumed to be $L=r_\mathrm{em}$.

\subsection{Losses}
\label{ssec:loss}
The FR-0 environment is radiation-dominated, meaning losses due to particle-particle interactions are negligible. In such a source region, charged nuclei will suffer from synchrotron radiation, electromagnetic pair production, photo-disintegration, and photo-meson production. Here, we briefly report the loss time scales and refer the interested reader to \cite{Merten2021} and references therein for more details.

\paragraph{Synchrotron:}
The average loss time scale for an isotropic ultra-relativistic ($v=c$) nuclei ensemble due to synchrotron radiation is 
\begin{align}
    \tau_\mathrm{loss}^\mathrm{synch} = \frac{3\,c}{4\,\sigma_\mathrm{T}\,u_\mathrm{mag}} \frac{m_\mathrm{N}^3}{m_\mathrm{e}^2} \frac{1}{Z^4 \gamma} \quad . \label{eq:SynchLoss}
\end{align}
Here, $\sigma_\mathrm{T}$ is the Thomson cross-section, $u_\mathrm{mag}$ is the magnetic energy density, $Z$ the charge number, $m_e$ the electron mass, $m_\mathrm{N}$ the nucleus mass, and $\gamma$ is the particle's Lorentz factor. 

\paragraph{Bethe-Heitler pair production:}
Assuming an isotropic photon target photon field, (see Section \ref{ssec:target}) the electromagnetic loss time scale is given by 
\begin{align}
    \left(\tau_\mathrm{loss}^\mathrm{BH}\right) ^{-1} = \gamma^{-1}\,\alpha_\mathrm{S}\,r_0^2\,c\, Z^2 \frac{m_e}{m_\mathrm{N}}\,\int_2^\infty \mathrm{d}\kappa\,n_\gamma\left(\frac{\kappa}{2\gamma_\mathrm{N}}\right)\frac{\phi(\kappa)}{\kappa^2} \quad \label{eq:BHLL} .
\end{align}
Here,~$\alpha_\mathrm{S}$ is the fine-structure constant, $r_0$ is the classical electron radius, and~$Z$ is the nucleus charge number. The function~$\phi(\kappa)$ is a semi-analytical description of the cross-section integral~(see Eq. 3.12 of \cite{Chodorowski1992}) where the photon energy in the nucleus rest frame is parametrized as $\kappa=2\gamma_\mathrm{N}\epsilon/(m_e c)$.

\paragraph{Photo-disintegration (PDI) and Photo-pion Production (PPP):}
The loss time scale for the two photo-hadronic losses is given by
\begin{align}
   \left(\tau_\mathrm{loss}^\mathrm{PDI/PPP}\right)^{-1} = \frac{c}{2\,\gamma^2_\mathrm{N}} \int_0^\infty \mathrm{d} \epsilon \frac{n_\gamma(\epsilon)}{\epsilon^2} \int_{\epsilon_\mathrm{r, thr}} \mathrm{d}\epsilon_\mathrm{r}\,\epsilon_\mathrm{r}\sigma_{\mathrm{N}\gamma}(\epsilon_\mathrm{r})\,\frac{\langle E \rangle}{E} \quad , \label{eq:Interactionrate}
\end{align}
where $\sigma_\mathrm{N\gamma}$ is the cross-section, $\epsilon_\mathrm{r}$ and $\epsilon$ are the photon energy in the nucleus and laboratory rest frame, respectively, and $\epsilon_\mathrm{r, thr}$ is the production threshold. The inelasticity is calculated as the weighted mean over the photo-disintegration decay channels. The photo-pion production of heavy CR nuclei ($A>1$) is approximated by simple superposition (see \cite{Merten2021} for the details).

\begin{figure}[htbp]
    \centering
    \includegraphics[width=.7\textwidth]{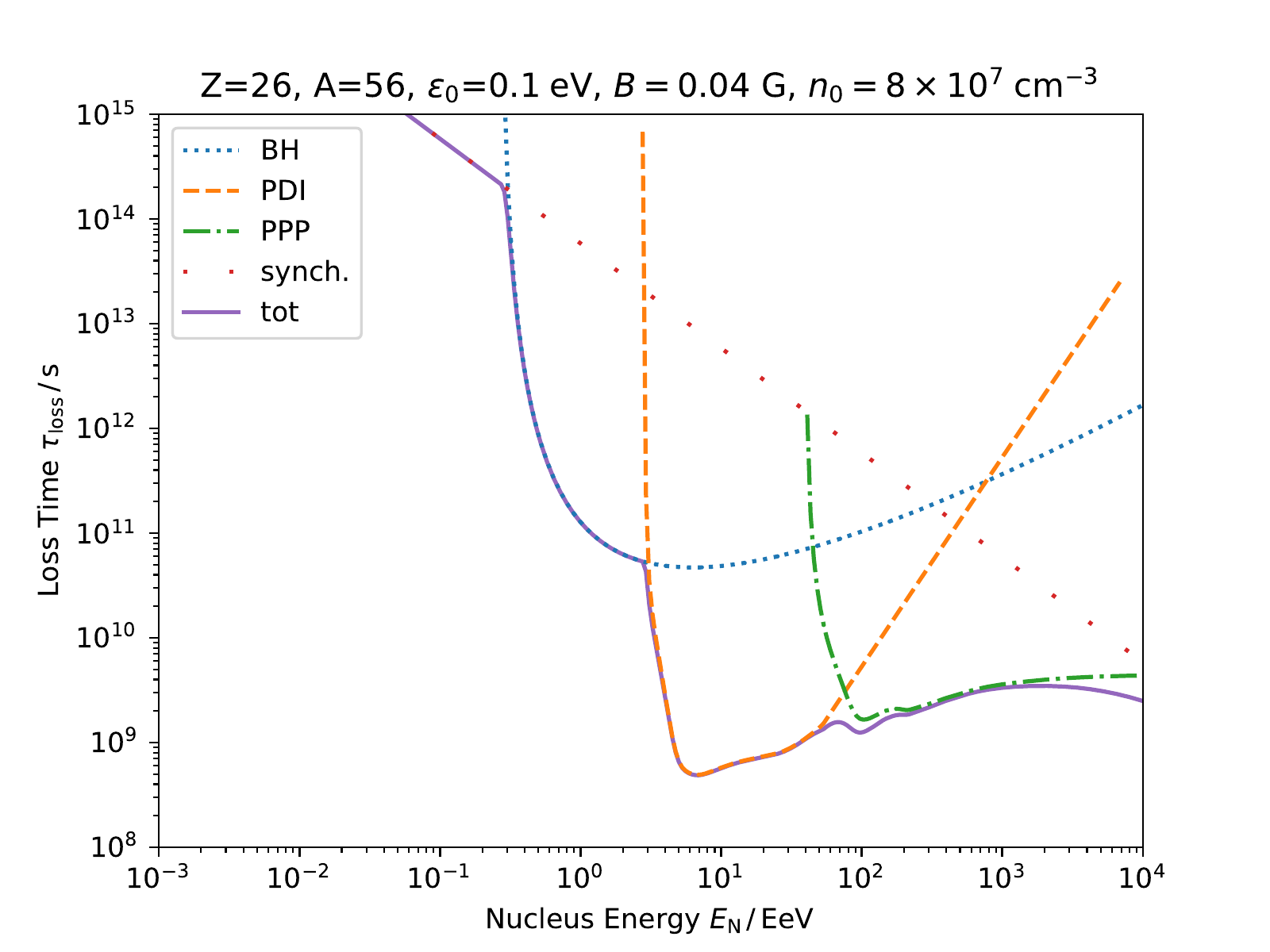}
    \caption{Shown are loss time scales for iron. The blue dotted line~(BH) indicates Bethe-Heitler pair production, which is combined with photo-disintegration~(PDI, orange-dashed), photo-pion-production~(PPP, green-dash-dotted), and synchrotron radiation (synch., red-sparse-dotted) to give the total loss length scale~$L_\mathrm{loss}$~(violet-solid). The parameters approximate the source environment in FR~0s.}
    \label{fig:loss}
\end{figure}

Figure \ref{fig:loss} shows the exemplary loss length for iron nuclei in a mono-energetic ($\epsilon=0.1\,\mathrm{eV}$) photon field with a target density of $n=8\times 10^7\,\mathrm{cm^{-3}}$ and a magnetic field strength of $B=0.04\,\mathrm{G}$. 

\section{FR-0 source environment}
\label{sec:Method}
Detailed observations of FR-0 source regions are not yet available. Based on the premise of assessing the UHECR acceleration in FR-0 jets, we assume a spherical acceleration region with radius $r_\mathrm{em}$ and a magnetic field strength $B$, where the product of the two parameters should be maximal without violating the maximal jet luminosity. The available information on the SED is compiled into a semi-analytical description for the target photon field, allowing to fix additional parameters, such as the jet Lorentz factor $\Gamma_\mathrm{j}$.

\subsection{Target Photon Field}
\label{ssec:target}
We use a simple approach based on a single zone synchrotron-self-Compton model and a host galaxy contribution to describe the target photon field. As the data is sparse, the SED was not fitted but only checked to not overshoot any observational constraints. Several different SED models were tested, but here only the most promising is shown (see \cite{Merten2021} for other models). The parameters of the SSC model are given in Table \ref{tab:SSCparams}. The host galaxy contribution is modeled with a blackbody spectrum and de-Vaucouleur radial profile, as shown in Table \ref{tab:bb}.
\begin{table}
	\centering
    \caption{Parameters of the SSC model. The following additional parameters are set:~$\gamma'_\mathrm{min}=100$,~$p_1=2$,~$p_2=3$, and~$\theta = 20 ^\circ$. The luminosity is given in logarithmic units~$\log_{10}(L_i/(\mathrm{erg}\,\mathrm{s}^{-1}))$, where a luminosity ratio between protons and electrons of~$\xi=10$ is assumed.}
    \label{tab:SSCparams}
    \begin{tabular}{rrrrrrrrrrr}
        \hline
        \multicolumn{1}{l}{r$_\mathrm{em}'/\mathrm{cm}$} & \multicolumn{1}{l}{$B'/\mathrm{G}$} &  \multicolumn{1}{l}{$\Gamma$} & \multicolumn{1}{l}{$\gamma_\mathrm{cut}'$} &  \multicolumn{1}{l}{$\gamma_\mathrm{max}'$} &  \multicolumn{1}{l}{$u_e'/(\mathrm{erg}\,\mathrm{cm}^{-3})$} &  \multicolumn{1}{l}{$u_\gamma'/(\mathrm{erg}\,\mathrm{cm}^{-3})$} &   \multicolumn{1}{l}{$L_\mathrm{equ}$} & \multicolumn{1}{l}{$L_\mathrm{CR}$}\\
        
        $5\times 10^{17}$ & 0.04 & 2 & $1\times 10^4$ & $3\times 10^4$ &  $1.90 \times 10^{-5}$ &  $3.39\times 10^{-6}$  & 43.39 & 43.72 \\

    \hline   
    \end{tabular}
\end{table}
\begin{table}[htbp]
    \centering
    \caption{Parameters of the host galaxy model. The radial profile follows the de Vaucoleur model with half-light radius $R_\mathrm{e}$ and scaling exponent $m$. The spectral distribution is given by a diluted blackbody with temperature $T$ and energy density $\epsilon_\mathrm{dil}$ resulting in a total luminosity $L_\mathrm{tot}$.}
    \label{tab:bb}
    \begin{tabular}{lrrrrr}
        \hline
        & \multicolumn{1}{l}{$L_\mathrm{tot}/(\mathrm{erg}\,\mathrm{s}^{-1})$} & \multicolumn{1}{l}{$T/\mathrm{K}$} & \multicolumn{1}{l}{$\epsilon_\mathrm{dil}/(\mathrm{cm}^3\,\mathrm{erg}^{-1})$} & \multicolumn{1}{l}{$R_\mathrm{e}/\mathrm{kpc}$} & \multicolumn{1}{l}{$m$} \\

        Blackbody &  $ 6.4\times 10^{44}$ & 4801 & 0.29 & 1 & 4 \\
        \hline
        
        \end{tabular}
\end{table}

\begin{figure}[htbp]
    \centering
    \includegraphics[width =.75\textwidth]{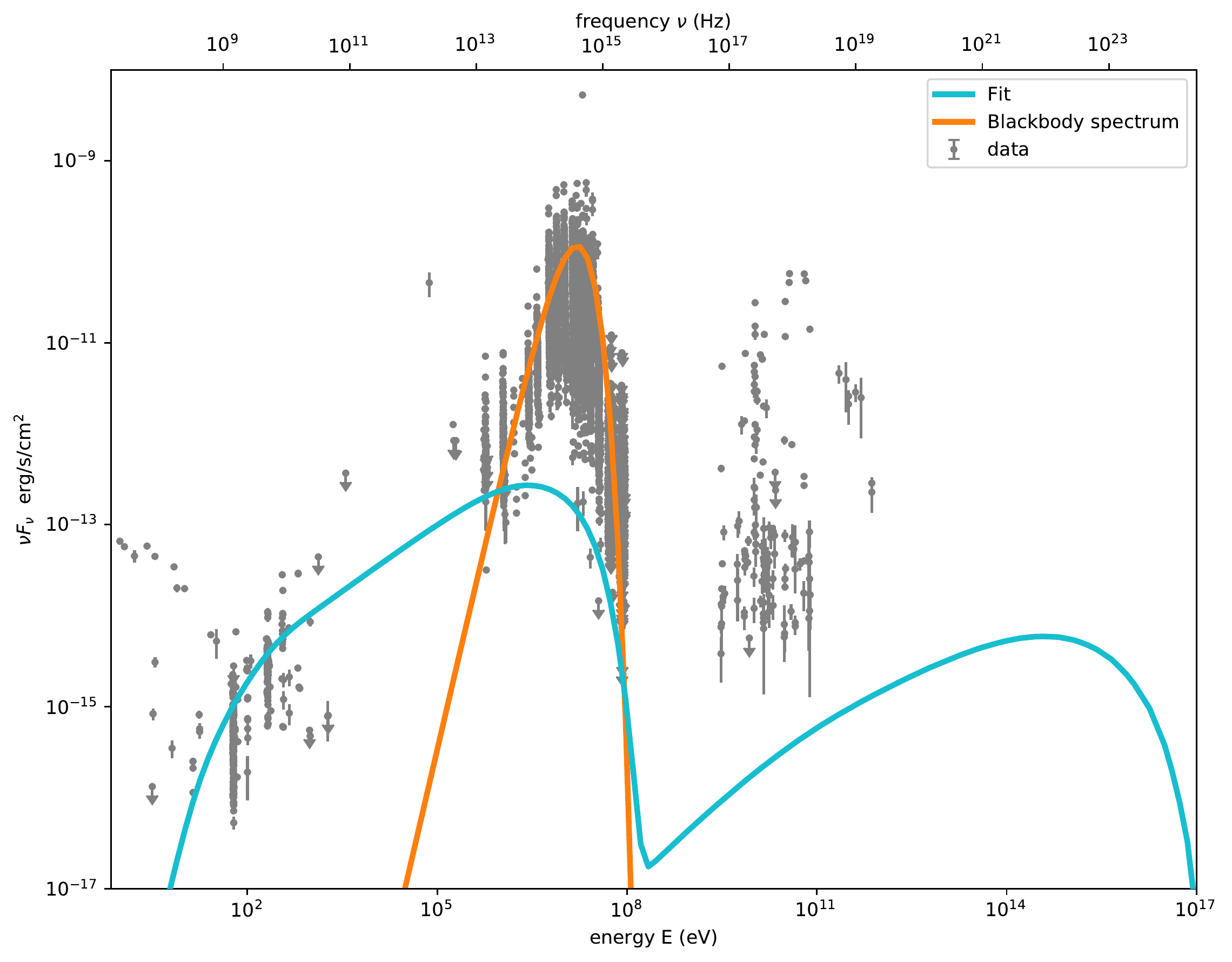}
    \caption{The combined flux spectrum of the~114 FR~0s (gray dots) and the derived target photon field approximation are shown. The host galaxy contribution (orange) is limited to a small frequency band. Furthermore, the SSC jet component (cyan) leaves room for a hadronic contribution at higher energies.} 
    \label{fig:average_SED}
\end{figure}

Figure \ref{fig:average_SED} shows the observed SED data of all 114 FR-0s and the target photon field that leads to the most promising UHECR results. The high-energy data are not fully explained by this SSC model leaving room for hadronic contributions.

\begin{figure}[htbp]
    \centering
    \begin{minipage}{.49\linewidth}
    \includegraphics[width=\linewidth]{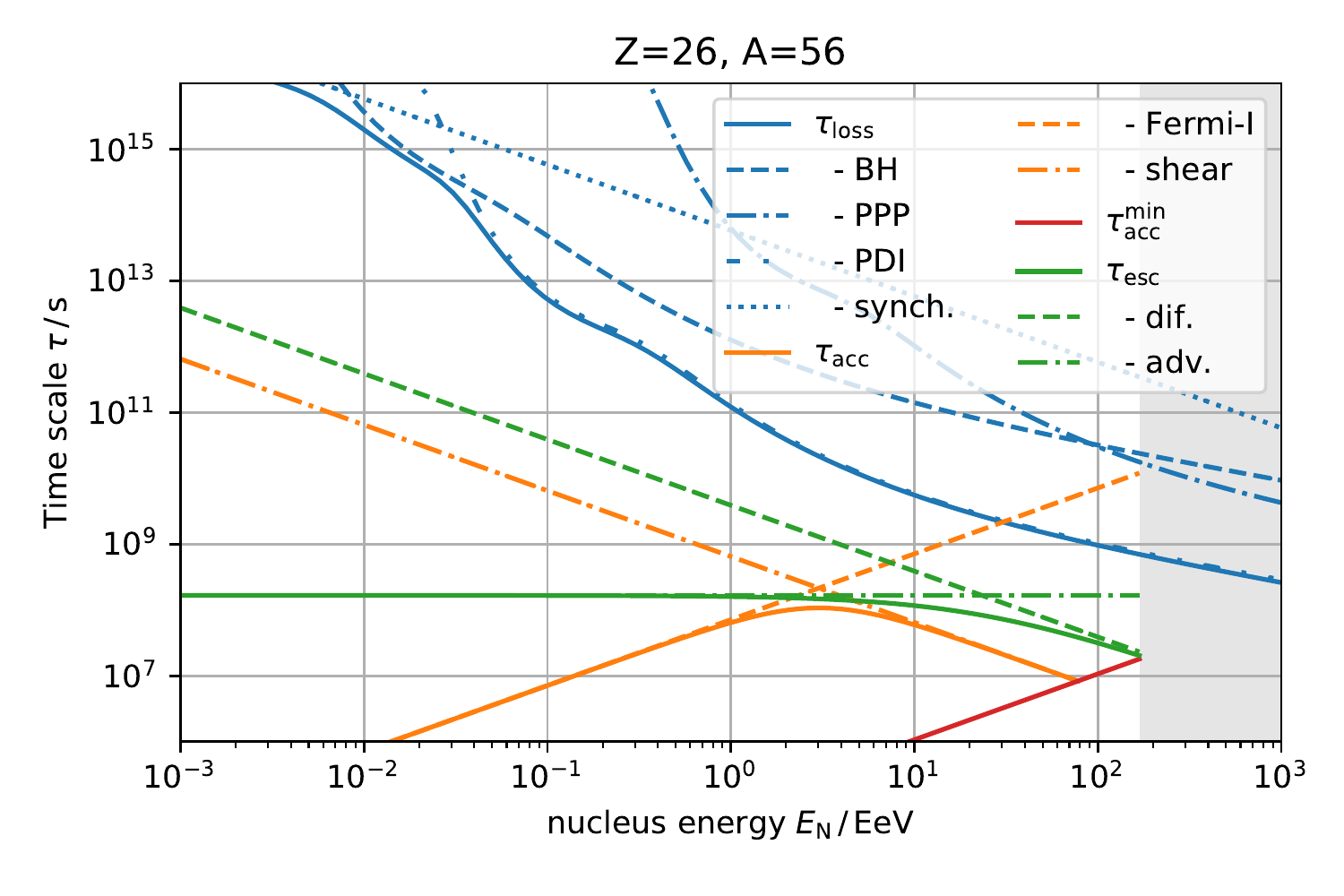}
    \end{minipage}%
    \begin{minipage}{.49\linewidth}
    \includegraphics[width=\linewidth]{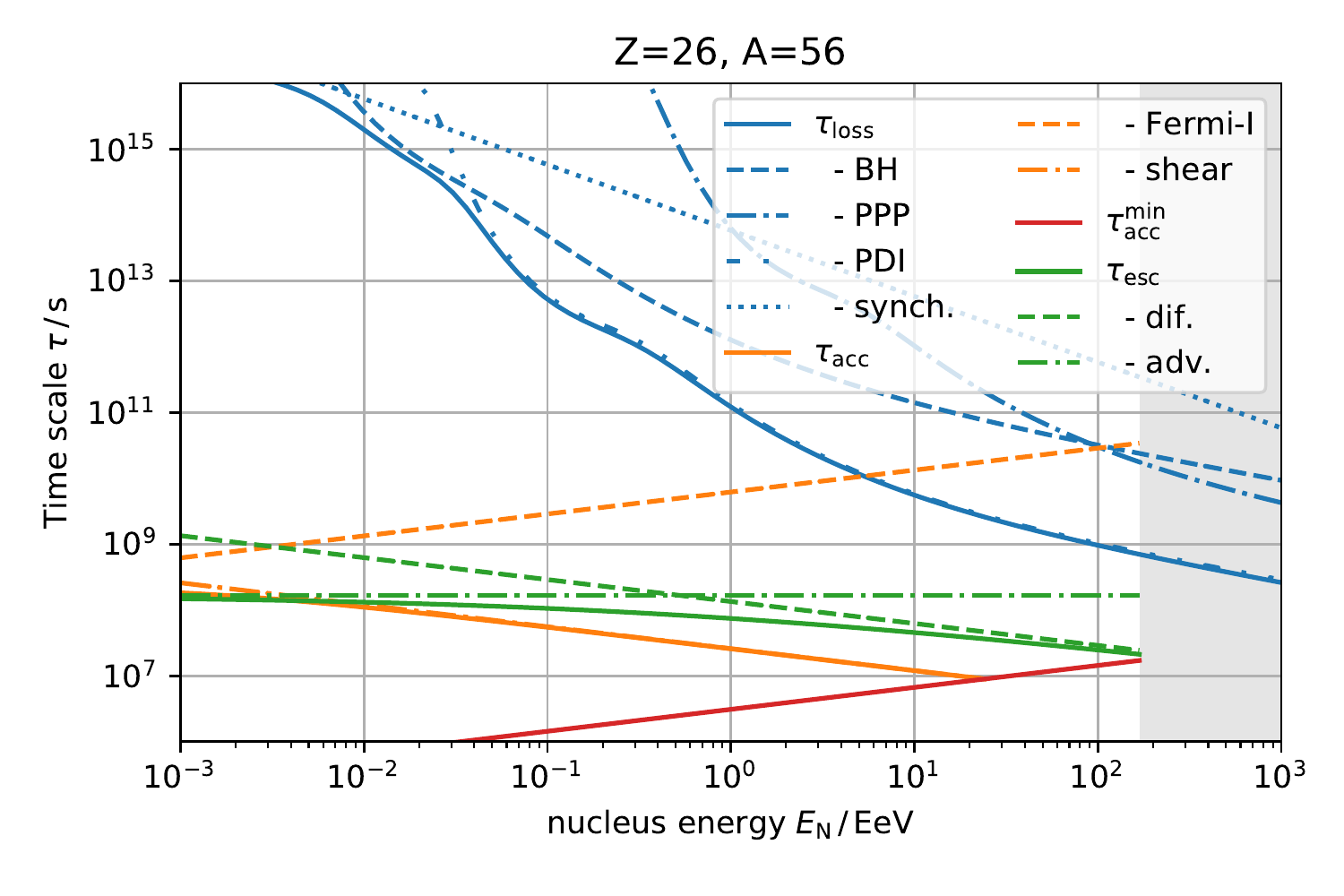}
    \end{minipage}
    \caption{Shown are the relevant time scales for iron. The left panel shows Bohm diffusion and the right panel shows Kolmogorov diffusion. Losses are shown in blue (compare to Figure \ref{fig:loss}), diffusive (advective) escape is shown in dashed (and dashed-dotted) green, and acceleration is shown as dashed (and dashed-dotted) orange lines for Fermi and gradual shear acceleration. The maximal acceleration scenario is shown as a solid red line, and the gray shaded area refers to energies above the Hillas criterion.}
    \label{fig:lengthsiron}
\end{figure}

Based on this target photon field, the time scale of all relevant processes (acceleration, escape, and loss) are shown in Figure \ref{fig:lengthsiron} for an iron nucleus. From this figure, it is immediately apparent that losses (blue lines) are not expected to dominate the particle transport. Furthermore, a scenario solely based on pure Fermi acceleration (orange dashed line) cannot reach the highest energies, as the escape process (solid green line) becomes more efficient around~$\sim 10$ EeV (Bohm diffusion-left panel), or even at~$\sim $PeV for Kolmogorov diffusion (right panel).

\subsection{Acceleration Probability and Maximal energy}
\label{ssec:pacc}
We define the acceleration probability, to determine the maximal energy for each tracer element and source configuration, as:
\begin{align}
    P_\mathrm{acc}(E) = \frac{\sum_i {\tau_{\mathrm{acc}, i}}^{-1}(E)}{\sum_i {\tau_{\mathrm{acc}, i}}^{-1}(E) + \sum_j{\tau_{\mathrm{esc}, j}}^{-1}(E) + \sum_k {\tau_{\mathrm{loss}, k}}^{-1}(E)} \quad .\label{eq:Pacc} 
\end{align}
Based on Equation~\ref{eq:Pacc} we define the maximum energy as the energy where the acceleration probability is halved $P_\mathrm{acc}(E_\mathrm{max}) = 0.5$. Afterward, this value is compared to the maximal energy given by the Hillas criterion, and the smaller of the two is reported.

\begin{figure}[htbp]
    \centering
    \begin{minipage}{.49\linewidth}
    \includegraphics[width=\linewidth]{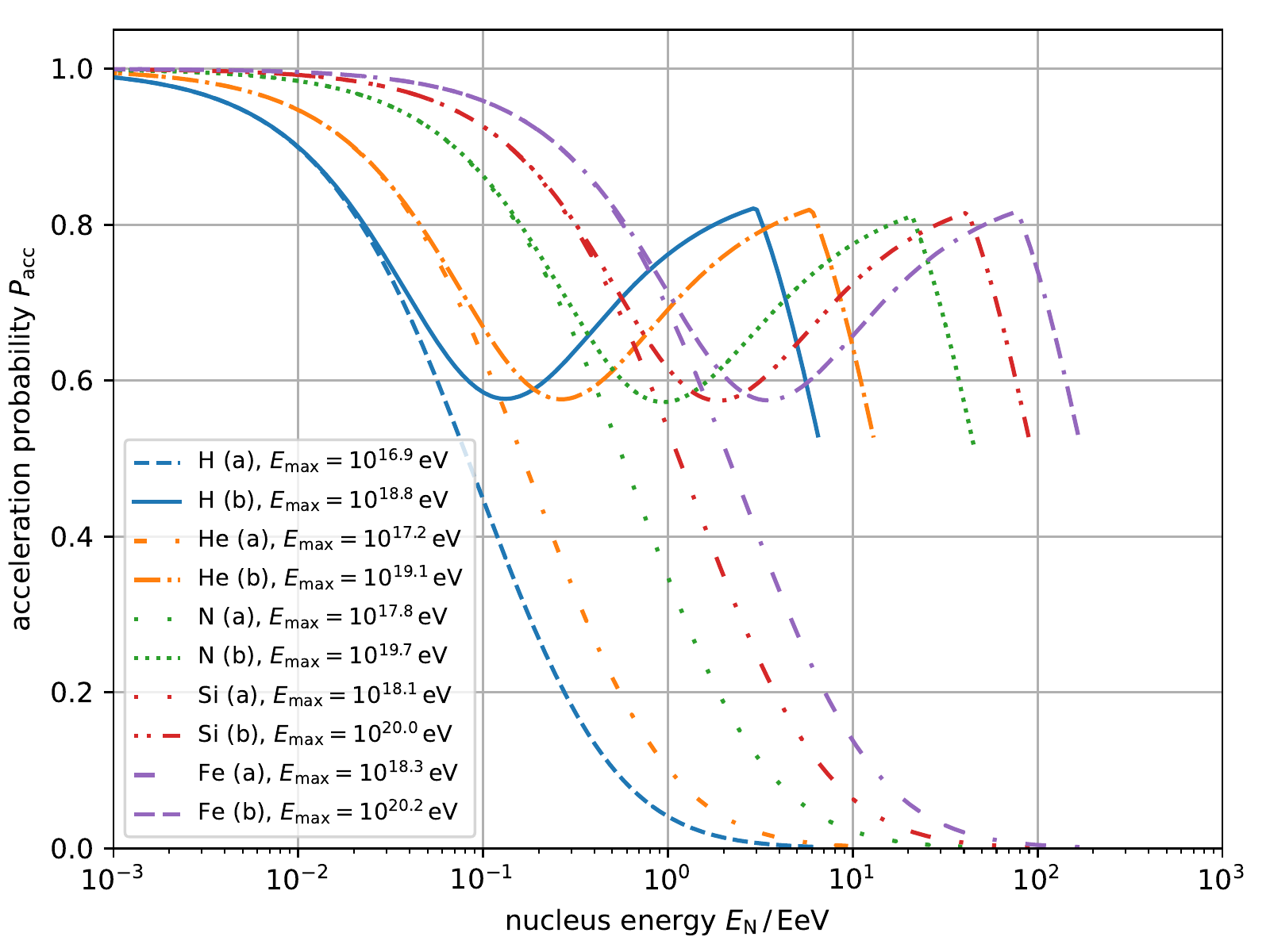}
    \end{minipage}%
    \begin{minipage}{.49\linewidth}
    \includegraphics[width=\linewidth]{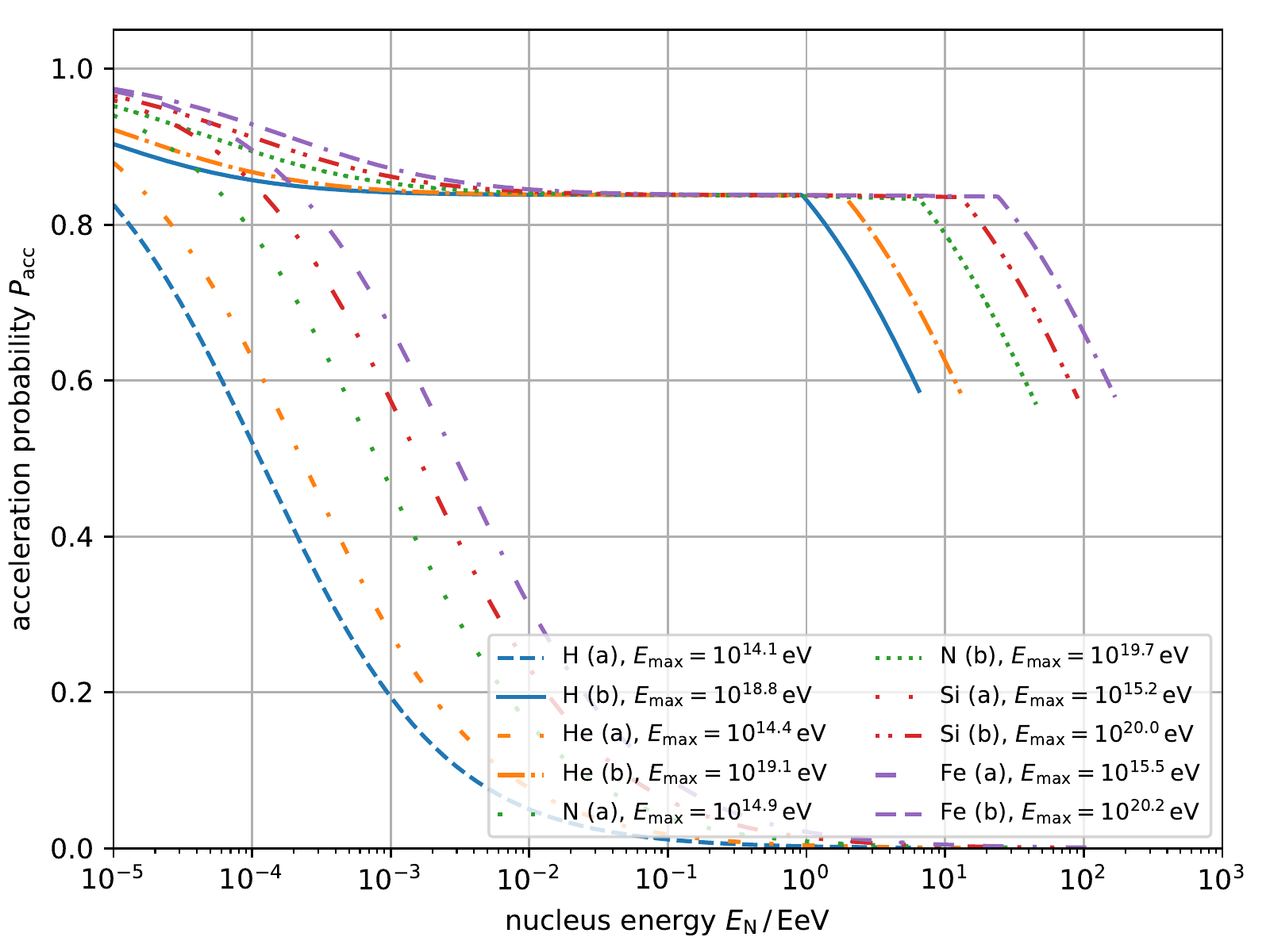}
    \end{minipage}
    \caption{Acceleration probability assuming Bohm (Kolmogorov) diffusion is shown in left (right) panel. A pure Fermi-1 order scenario is indicated by \textit{(a)}, whereas the more optimistic hybrid scenario is indicated with \textit{(b)}. The hard break in the hybrid scenario refers to the upper limit of efficient gradual shear acceleration.}
    \label{fig:pacc_iron}
\end{figure}

Figure~\ref{fig:pacc_iron} shows the acceleration probability for all elements calculated for both Bohm (left panel) and Kolmogorov diffusion scenarios. 
The maximal energies for the examined source model are shown in the legend of in Figure \ref{fig:pacc_iron}. The mean rigidity over all tracer elements in case of pure Fermi-1 acceleration and Bohm (Kolmogorov) diffusion is $\bar{R}_\mathrm{Bohm, Fermi}=10^{16.91\pm 0.03}\,\mathrm{V}$ ($\bar{R}_\mathrm{Kolm., Fermi}=10^{14.08\pm 0.02}\,\mathrm{V}$ is much smaller than for the hybrid scenario with $\bar{R}_\mathrm{Hybr.}=10^{18.81\pm 0.03}\,\mathrm{V}$, where the maximal energy is independent of the turbulence model.

\section{Conclusions} 
\label{sec:outlook}

The acceleration, escape, and loss time scales of an average FR-0 radio galaxy were examined by compiling the available SED into a two-component target photon field consisting of an internal jet component (SSC model) and an external host galaxy contribution. Combining all relevant timescales into the acceleration probability $P_\mathrm{acc}$ allowed a realistic assessment of the chances for UHECRs acceleration in these numerous sources. We conclude:
\begin{enumerate}
    \item Independent of the turbulence model, acceleration up to the highest observed energies is possible if a hybrid acceleration --- the combination of pre-acceleration, e.g., Fermi-1 order, with gradual shear acceleration --- is realized. These acceleration scenarios on their own cannot contribute to the UHECR flux.
    
    \item The exact shock speed $u_\mathrm{s}$ does not have a significant influence on the maximal energy, however, it determines the transition between Fermi-1 and gradual shear acceleration.
    
    \item Although FR-0 galaxies are comparatively radiatively weak sources, we expect significant deviations from an unbroken power law for the ejected cosmic-ray spectra due to energy-dependent losses.
\end{enumerate}

In summary, FR-0 radio galaxies represent an interesting source class for UHECR acceleration especially because of their large number density and rather isotropic distribution.

\section*{Acknowledgments}
This work acknowledges financial support from the Austrian Science Fund (FWF) under grant agreement number I 4144-N27 and from the Slovenian Research Agency - ARRS~(project no. N1-0111). M.B. has for this project received funding from the European Union’s Horizon 2020 research and innovation program under the Marie Sklodowska-Curie grant agreement No 847476. The views and opinions expressed herein do not necessarily reflect those of the European Commission. G.B.\ acknowledges financial support from the State Agency for Research of the Spanish MCIU through the "Center of Excellence Severo Ochoa" award to the Instituto de Astrofísica de Andalucía (SEV-2017-0709) and from the Spanish "Ministerio de Ciencia e Innovaci\'on” (MICINN) through grant PID2019-107847RB-C44.

This research has made use of the NASA/IPAC Extragalactic Database (NED), which is funded by the National Aeronautics and Space Administration (NASA) and operated by the California Institute of Technology (CIT). Furthermore, part of this work is based on archival data, software, or online services provided by the Space Science Data Center (SSDC) via ASI.

This work benefited from the following software: CRPropa~\citep{crpropa30, crpropa31}, NumPy \citep{numpy}, Matplotlib \citep{matplotlib}, pandas \citep{pandas}, SOPHIA \cite{sophia}, jupyter notebooks \cite{ipython}, AGN SED tool \citep{Massaro2006, Tramacere2009, Tramacere2011}.

\bibliography{literature}{}
\bibliographystyle{elsarticle-num-names}

\end{document}